%% file: main.tex
\documentclass[conference]{IEEEtran}

\usepackage{cite}
\usepackage{booktabs}
\usepackage{amsmath}
\usepackage{amssymb}
\usepackage{graphicx}
\usepackage{xcolor}
\usepackage{textcomp}
\usepackage{pifont}
\usepackage{url}
\usepackage{subcaption}

\newcommand{\cmark}{\ding{51}}
\newcommand{\xmark}{\texttimes}
\newcommand{\pmark}{\ensuremath{\triangle}}

\graphicspath{{./images/}}

\title{Modeling Edge-to-Cloud Offloading Workloads for Autonomous Vehicles}

\author{
\IEEEauthorblockN{Longkun Li\IEEEauthorrefmark{1} and Evangelos Pournaras\IEEEauthorrefmark{1}\IEEEauthorrefmark{2}}
\IEEEauthorblockA{\IEEEauthorrefmark{1}\textit{School of Computer Science, University of Leeds}, Leeds, UK\\
\IEEEauthorrefmark{2}\textit{School of Energy Systems, LUT University}, 53850 Lappeenranta, Finland\\
\{cmhx9984, e.pournaras\}@leeds.ac.uk}
}

\begin{document}

\maketitle

\begin{abstract}
Autonomous vehicles often need to upload high volumes of diverse data within the edge-to-cloud continuum: telemetry, sensor data for offline training, and observations used to maintain high definition (HD) maps. We introduce a workload generator that represents each process for individual vehicles and aggregates the resulting traffic for a fleet and its wireless access points. Using Munich city as a real-world case study, we combine SUMO vehicle activity, sensor parameters, and hourly weights derived from recorded road accidents. The studies varying the timing and number of training data selections, model parameters, and transfer deadlines show that, with a two-hour deadline, earliest deadline first scheduling reduces the peak workload by 8.3\% without discarding data. At the same total capacity, demand-based allocation reduces the median unserved load from 28.4\% under equal allocation to 0\% across five rolling evaluations. The generator provides profound reproducible inputs for capacity, scheduling, and placement experiments.
\end{abstract}

\begin{IEEEkeywords}
Autonomous vehicles, vehicle-to-everything communications, edge computing, edge-to-cloud offloading, workload modeling, capacity planning.
\end{IEEEkeywords}

\input{sections/01_introduction}
\input{sections/02_gap_motivation}
\input{sections/03_workload_model}
\input{sections/04_workload_instantiation}
\input{sections/05_conclusion}

\section*{Acknowledgment}
This research is supported by a UKRI Future Leaders Fellowship (MR/W009560/1) through the project \textit{Digitally Assisted Collective Governance of Smart City Commons--ARTIO}. The authors would also like to thank Chuhao Qin and Rabiya Khalid for their insights and feedback.

\bibliographystyle{IEEEtran}
\bibliography{references}

\end{document}

%% file: sections/01_introduction.tex
\section{Introduction}

Autonomous vehicles upload data to roadside, edge, and cloud infrastructure, including operational telemetry, selected sensor data for offline training, and observations for maintaining high definition (HD) maps~\cite{liu2019edge}. Perception and control remain onboard. The workload model can also provide inputs for future studies of whether communication and processing delays leave enough time for emergency braking. The resulting traffic is an important input to studies of vehicle-to-everything access and edge capacity.

Cameras, LiDAR, and radar can generate substantial raw data. Standards for advanced driving and extended sensors define service requirements~\cite{3gpp2020v2xrequirements} and identify multi-access edge computing as support for vehicle-to-everything services~\cite{etsi2018mecv2x}. Onboard filtering reduces these streams to outputs selected by each application~\cite{tomei2019sensor}; workload inputs should therefore reflect what is uploaded, when, and where to support rigorous experimentation.

Telemetry contributes persistent load. Sensor data for training are selected when vehicles encounter safety-related road events, while map observations arise at particular road locations~\cite{zhang2021real}. These applications differ in rate, size, timing, and location. A constant per-vehicle rate can match the total traffic generated by the model over 24 hours, but it cannot represent hourly changes caused by application timing.

Research on vehicular edge computing studies scheduling and resource allocation using tasks with fixed sizes, stationary arrivals, or data rates independent of road context~\cite{wang2020online}. These assumptions support controlled comparisons of algorithms, but they do not model the demand produced by autonomous vehicle applications.

We study how telemetry, training data, and map observations affect data traffic generated for vehicle-to-everything access networks as well as edge and cloud infrastructure. Specifically, we ask three questions. \textbf{RQ1:} How do the timing of application uploads and variation in hourly training data selection affect fleet peak workload relative to an assumed constant per-vehicle rate? \textbf{RQ2:} How do workload parameters and bounded transfer delay affect peak demand? \textbf{RQ3:} How does capacity allocation based on earlier demand affect unserved load across access points under the same total capacity?

The proposed model represents each process and aggregates demand for vehicles, the fleet, and each access point. A real-world case study in Munich provides a novel combination of realistic vehicle activity, a mean rate of training data generation, and relative hourly safety weights derived from recorded road accidents and simulated concurrent vehicle counts.

\noindent\textbf{Contributions.}
The contributions are: (1) a parameterized generator for telemetry, training data, and map observations, with demand reported for vehicles, the fleet, and each access point; (2) a reproducible method that controls the mean rate of training data generation independently of its hourly distribution; (3) new experimental insights into how hourly weights, stochastic hourly counts, model parameters, and transfer deadlines affect peak workload; and (4) evidence that demand-based capacity allocation reduces unserved load across access points under the same total capacity. The implementation and traces are publicly available.\footnote{\url{https://github.com/TDI-Lab/av-edge-workload-model}}

%% file: sections/02_gap_motivation.tex
\section{Research Gap and Motivation}
\label{sec:gap_motivation}

Vehicular edge research includes offloading, scheduling, resource allocation, and server placement~\cite{liu2019edge,wang2020online,chen2023mobility}. These studies usually take task sizes and arrival rates as inputs. Therefore, they do not model when autonomous vehicle applications upload data or how much each application contributes.

Mobility-aware resource allocation and decentralized service placement use vehicle movement to estimate where computing demand occurs~\cite{nezami2025computing,nezami2021decentralized}. Decentralized traffic optimization also models vehicle movement across a city, but it does not derive the traffic uploaded by vehicle applications~\cite{gerostathopoulos2019trapped}. OpenVDAP and CoPace process selected vehicle data or tasks~\cite{zhang2018openvdap,tian2021copace}, while CADET supports cooperative autonomy experiments~\cite{sharma2026cadet}. These systems take selected data or task arrivals as inputs; they do not derive the resulting traffic from what vehicle applications collect and upload.

Other studies address sensor data reduction~\cite{tomei2019sensor}, HD map maintenance~\cite{zhang2021real}, or measured edge workloads~\cite{kim2024edge}. Each addresses only part of the generated traffic. None combines telemetry, selected training data, and HD map observations with vehicle activity and location in one workload model.

Table~\ref{tab:related_gap} compares six representative studies. Upload requirements specify what data a vehicle sends and the size of each transmission. Mobility awareness means that demand follows vehicle activity or movement. Spatial awareness means that demand is associated with location, and infrastructure evaluation covers capacity, scheduling, or placement experiments.

\begin{table*}[t]
\centering
\small
\setlength{\tabcolsep}{5pt}
\caption{Comparison of support for requirements governing data upload and for evaluating infrastructure. \cmark: explicit support; \pmark: partial support; \xmark: not modeled.}
\label{tab:related_gap}
\begin{tabular}{lcccc}
\toprule
\textbf{Work} & \textbf{Upload Requirements} & \textbf{Mobility Awareness} & \textbf{Spatial Awareness} & \textbf{Infrastructure Evaluation} \\
\midrule
Zhang et al. (2018)~\cite{zhang2018openvdap} & \xmark & \xmark & \cmark & \pmark \\
Nezami et al. (2025)~\cite{nezami2025computing} & \pmark & \cmark & \cmark & \cmark \\
Wang et al. (2020)~\cite{wang2020online} & \xmark & \cmark & \xmark & \pmark \\
Tomei et al. (2019)~\cite{tomei2019sensor} & \pmark & \xmark & \xmark & \xmark \\
Zhang et al. (2021)~\cite{zhang2021real} & \cmark & \pmark & \cmark & \xmark \\
Kim and Gangadharan (2024)~\cite{kim2024edge} & \pmark & \cmark & \cmark & \cmark \\
\midrule
\textbf{This work} & \cmark & \cmark & \cmark & \cmark \\
\bottomrule
\end{tabular}
\end{table*}

\begin{figure*}[t]
\centering
\begin{subfigure}[b]{0.32\textwidth}
\centering
\includegraphics[height=3.7cm,keepaspectratio]{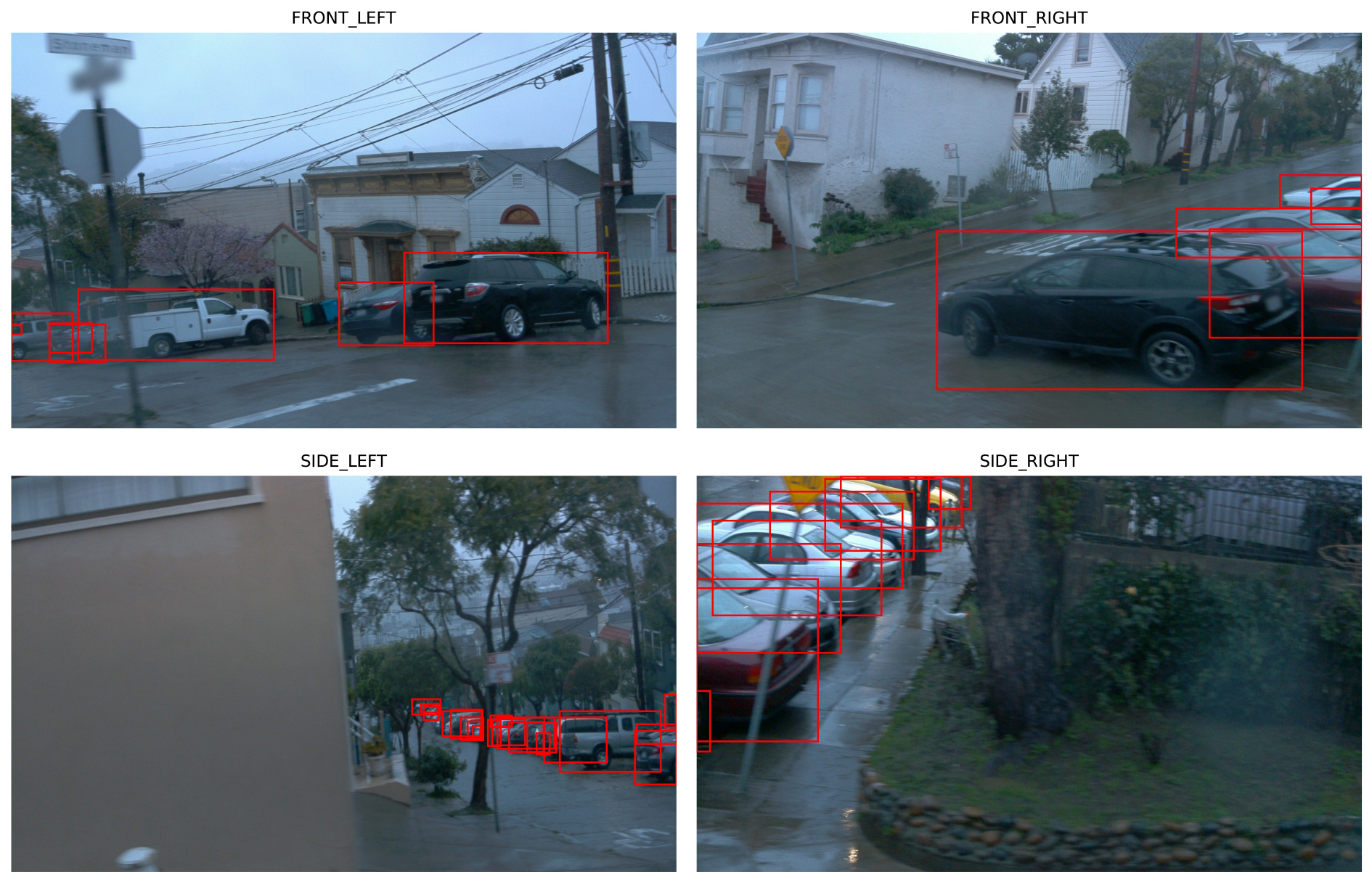}
\caption{Multiple camera views}
\end{subfigure}\hfill
\begin{subfigure}[b]{0.32\textwidth}
\centering
\includegraphics[height=3.7cm,keepaspectratio]{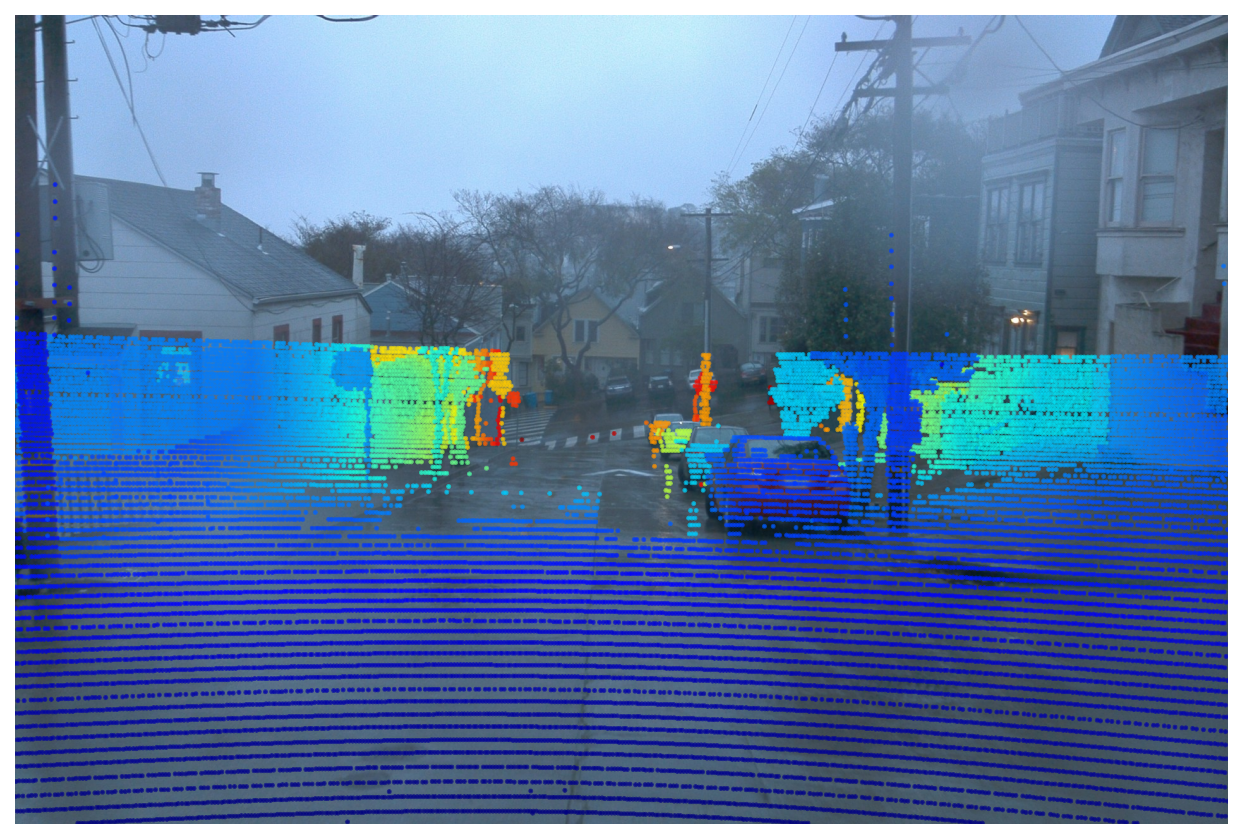}
\caption{Camera and LiDAR fusion}
\end{subfigure}\hfill
\begin{subfigure}[b]{0.32\textwidth}
\centering
\includegraphics[height=3.7cm,keepaspectratio]{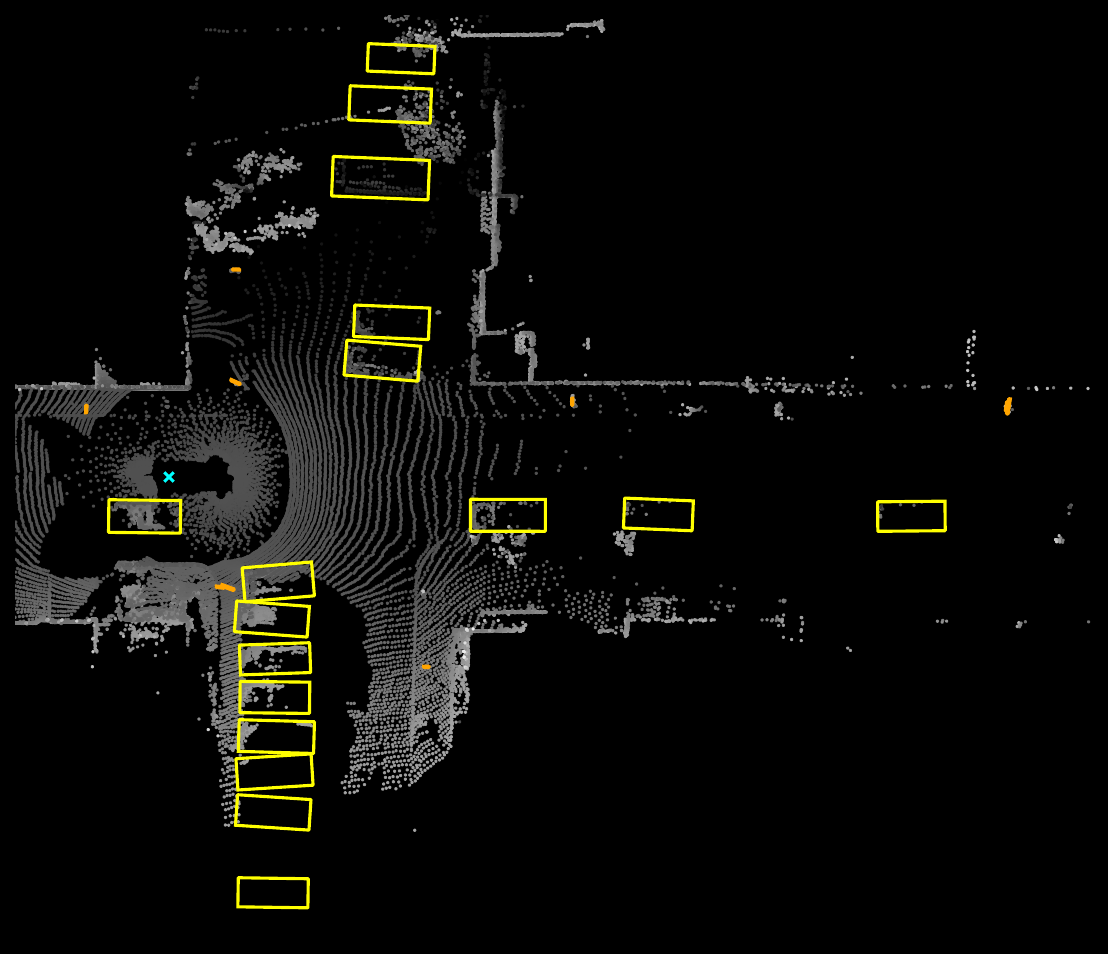}
\caption{LiDAR point cloud}
\end{subfigure}
\caption{Representative outputs from one Waymo Open Dataset frame: (a) multiple synchronized camera views; (b) LiDAR points projected onto a camera image; and (c) a LiDAR point cloud.}
\label{fig:waymo_multimodal}
\end{figure*}

Fig.~\ref{fig:waymo_multimodal} shows camera and LiDAR outputs from the Waymo Open Dataset~\cite{sun2020scalability}. These raw streams remain onboard for perception and control. Only data retained for telemetry, offline training, or HD map maintenance are included in the workload.

These studies lack a model that converts autonomous vehicle application requirements into data traffic before network scheduling or infrastructure capacity allocation. Telemetry, offline training, and map maintenance are used for advanced driving, extended sensing, and video-sharing services~\cite{3gpp2020v2xrequirements}.

%% file: sections/03_workload_model.tex
\section{Workload Modeling Framework}
\label{sec:workload}

The fleet workload is modeled as the sum of three application data flows: periodic telemetry, sensor data selected for offline training, and observations used for HD map maintenance. Only data selected for transmission are included in the workload.

\subsection{System Overview}

\begin{figure*}[t]
\centering
\includegraphics[width=\linewidth]{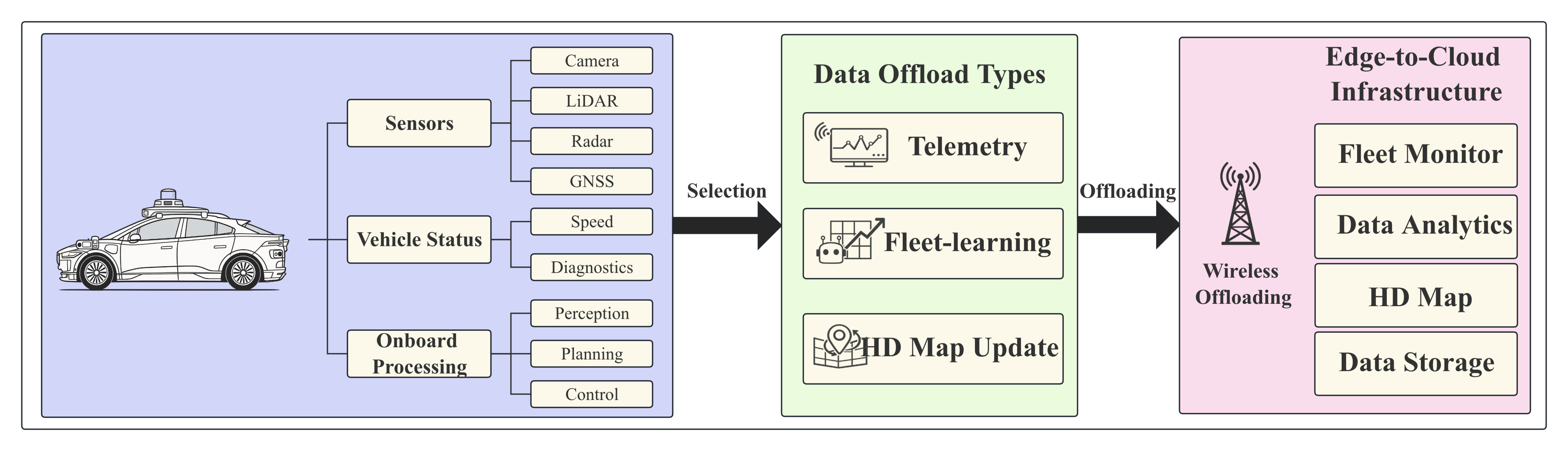}
\caption{Architecture for transferring data from vehicles to cloud services. Vehicles keep perception and control with strict latency requirements onboard and upload selected telemetry, training data, and HD map data through wireless access points or edge service areas.}
\label{fig:system_model}
\end{figure*}

We consider a fleet of autonomous vehicles:
\begin{equation}
\mathcal{V}=\{v_1,v_2,\ldots,v_N\},
\label{eq:fleet_set}
\end{equation}
operating in a road network. Let $\mathcal{V}(t)\subseteq\mathcal{V}$ denote the vehicles active at time $t$, and let $N(t)=|\mathcal{V}(t)|$. Inactive vehicles are excluded. Perception and control remain onboard, while selected data traverse the wireless network. In this model, an access point is the wireless attachment location used to aggregate uploaded data. It may represent a roadside unit, a cellular site, or the attachment point for an edge service area. Fig.~\ref{fig:system_model} shows the data path.

We model three representative categories of uploaded data:
\begin{itemize}
\item \textbf{Telemetry data.} Vehicles periodically transmit operational state, diagnostics, and compact summaries such as speed, localization status, and health indicators.
\item \textbf{Training data.} Vehicles may retain several seconds of camera, LiDAR, and radar data when a road event, such as a perception failure or system anomaly, triggers data collection. Offline training means that the uploaded recordings are used later to train models, not for real-time perception or control while the vehicle is driving.
\item \textbf{HD map data.} Vehicles may upload compact observations of a displaced road sign, altered lane geometry, or temporary construction after onboard processing identifies a possible change.
\end{itemize}

\subsection{Model Parameters}

Let $\mathcal{S}$ denote the set of onboard sensor types, such as camera, LiDAR, and radar. Data sizes are measured in bytes and rates in bytes per second.

For each sensor type $s\in\mathcal{S}$, we define:
\begin{itemize}
\item $n_s$: number of sensors of type $s$ installed on a vehicle;
\item $f_s$: sampling frequency, such as a camera frame rate;
\item $b_s$: mean encoded size in bytes of one measurement from sensor type $s$;
\item $p_s$: fraction of the raw sensor byte rate $n_sf_sb_s$ included in periodic telemetry.
\end{itemize}

The parameters for training data selected after road events are outlined below:
\begin{itemize}
\item $\lambda(t)$: mean rate at which one active vehicle retains a sensor recording after a road event triggers data collection; examples include perception uncertainty and system anomalies;
\item $T$: duration of the stored sensor window;
\item $a_s$: proportion of measurements from sensor stream $s$ retained onboard for offline training during a window of $T$ seconds; $a_s=1$ retains all measurements and $a_s=0.5$ retains half on average;
\item $c_s$: ratio of encoded upload bytes to the retained bytes from sensor stream $s$ after additional compression.
\end{itemize}
Lower values of $a_s$ or $c_s$ reduce traffic but may also reduce the detail retained for offline training.

The parameters for HD map data are outlined below:
\begin{itemize}
\item $u(t)$: rate at which an active vehicle produces compact map data items, such as records for lane boundaries or road signs;
\item $m$: number of map features in one item;
\item $b_M$: mean encoded size of one map feature;
\item $r_E(t)$: byte rate of supporting evidence attached to compact map records after onboard processing, such as a cropped image or selected point-cloud data; $r_E(t)=0$ represents map metadata only.
\end{itemize}

\subsection{Workload for One Vehicle}

For sensor type $s$, the data rate produced by its sensors is:
\begin{equation}
r_s=n_sf_sb_s.
\label{eq:raw_sensor_rate}
\end{equation}
For a camera, the product combines its count, frame rate, and encoded frame size. Only selected data contribute to offloading demand.

Compact summaries and operational state form the periodic telemetry rate:
\begin{equation}
r_T=\sum_{s\in\mathcal{S}}p_sr_s.
\label{eq:telemetry_rate}
\end{equation}
The factor $p_s$ represents the equivalent share carried by compact telemetry records; it does not imply that raw frames are continuously uploaded.

When a road event triggers data collection, the vehicle stores a short sensor window for offline training. The mean encoded data volume produced by one such event is:
\begin{equation}
B_L=T\sum_{s\in\mathcal{S}}a_sc_sr_s.
\label{eq:training_volume}
\end{equation}
Here, $T$ sets the window duration, $a_s$ determines how many measurements from stream $s$ are retained, and $c_s$ captures subsequent compression. The expected training data rate of one active vehicle is:
\begin{equation}
r_L(t)=\lambda(t)B_L.
\label{eq:training_rate}
\end{equation}

For HD map maintenance, the mean size of one compact map data item is:
\begin{equation}
B_M=mb_M.
\label{eq:map_item}
\end{equation}
For example, one item may contain several lane, sign, or road geometry features. Adding supporting evidence gives the expected map data rate:
\begin{equation}
r_M(t)=u(t)B_M+r_E(t).
\label{eq:map_rate}
\end{equation}
The first term represents compact map records, and the second represents supporting evidence attached after onboard processing. The expected offloading workload of one active vehicle is therefore:
\begin{equation}
r(t)=r_T+r_L(t)+r_M(t).
\label{eq:vehicle_rate}
\end{equation}

\subsection{Aggregation Across the Fleet and Access Points}

Because the base model applies the same configuration to all active vehicles, the total fleet workload is:
\begin{equation}
R(t)=N(t)r(t).
\label{eq:fleet_rate}
\end{equation}
For vehicles with different sensor configurations or collection policies, the implementation sums individual rates, providing the heterogeneous vehicle support reported in Table~\ref{tab:related_gap}.

Vehicles attached to different access points do not use the same local service capacity. Let $N_a(t)$ denote the number of active vehicles served by access point $a$ at time $t$. The workload generated by vehicles served by access point $a$ is:
\begin{equation}
R_a(t)=N_a(t)r(t).
\label{eq:location_rate}
\end{equation}
If each active vehicle is connected to exactly one access point, then $\sum_aN_a(t)=N(t)$ and $\sum_aR_a(t)=R(t)$. For heterogeneous vehicles, individual rates are summed at each access point. This converts vehicle activity into local demand for identifying spatial hotspots and testing capacity allocation. Mobility changes $N_a(t)$ as vehicles move, while collection requirements determine the workload of each vehicle.

%% file: sections/04_workload_instantiation.tex
\section{Workload Instantiation}
\label{sec:instantiation}

The mobility input is a 24-hour SUMO trace~\cite{krajzewicz2010traffic} derived from the Munich road network, with origin and destination demand used by Nezami et al.~\cite{nezami2025computing}. Let $N_h$ denote the mean number of concurrently active vehicles in hour $h$. The German Unfallatlas is a public database of road accidents involving personal injury~\cite{unfallatlas}. Its Munich records contain 6,940 accidents from 2023; let $A_h$ denote the number of recorded accidents in hour $h$. We use this hourly distribution as an observable proxy for when safety-related road events that trigger sensor recording may occur more frequently.

The accident records and mobility data have different absolute scales, so we do not compare their counts directly. Instead, we use them to derive a relative hourly weight. For each hour, we divide the historical accident count $A_h$ by the mean number of concurrent vehicles $N_h$ to account for differences in vehicle activity and then normalize the result:
\begin{equation}
q_h=\frac{A_h/N_h}{\sum_jA_j/\sum_jN_j},\qquad
\lambda(h)=\frac{\bar{\lambda}q_h}{3600}.
\label{eq:event_timing}
\end{equation}
The normalization removes the absolute scale of both data sets. The dimensionless value $q_h$ is neither an accident probability nor a simulated accident count; it controls only the relative hourly variation in training-data selection. Its vehicle-weighted daily mean is one. The parameter $\bar{\lambda}$ independently sets the mean number of training-data selections per active vehicle per hour. At the hourly resolution used in the evaluation, $N_h$, $\lambda(h)$, and $R(h)$ denote the mean concurrent vehicle count, the selection rate per active vehicle, and the mean fleet workload in hour $h$, respectively.

\begin{figure*}[t]
\centering
\includegraphics[width=\linewidth]{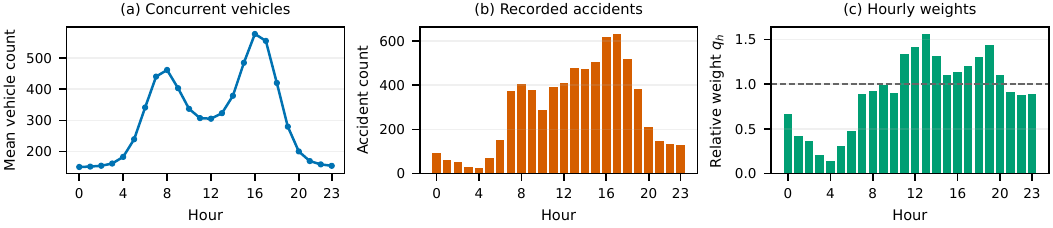}
\caption{Munich workload inputs: (a) mean concurrent vehicles; (b) recorded accidents; and (c) relative hourly weights.}
\label{fig:data_grounding}
\end{figure*}

Fig.~\ref{fig:data_grounding} shows the hourly vehicle counts $N_h$, accident counts $A_h$, and relative weights $q_h$.

\subsection{Scenario and Parameters}

The main configuration treats all vehicles in the SUMO trace as autonomous vehicles; a lower share can be represented by applying the workload model to the corresponding proportion of vehicles. We compare three workload representations. The \emph{telemetry control} contains only periodic telemetry traffic. The \emph{application model} contains telemetry, training data, and HD map data. The \emph{fixed-rate baseline} assigns every active vehicle the same constant rate while preserving the same total data volume over 24 hours. This rate is:
\begin{equation}
r_c=\frac{\sum_hR(h)}{\sum_hN_h}.
\label{eq:constant_rate}
\end{equation}
The fixed-rate baseline and application model therefore differ in timing, not transferred volume.

Table~\ref{tab:configuration} lists the main configuration. Sensor counts follow the sixth-generation Waymo Driver~\cite{waymo_driver_2024}, and the 10 Hz sampling rate follows the Waymo Open Dataset~\cite{sun2020scalability}. We set the encoded sizes and filtering parameters for this scenario, yielding 0.527 Mbps of telemetry and 18.6 MiB of training data per event. The encoded fraction of 0.1 represents a tradeoff between retained information and upload volume~\cite{tomei2019sensor}. Because public data do not report selection frequency, we use one selection per vehicle per hour and test rates from 0.25 to 4. We convert the 126 KB/km map evidence value to a data rate at 30 km/h~\cite{zhang2021real}; compact map records add 10 kB every 20 s. These rates lie within the ranges of enhanced vehicle-to-everything services~\cite{3gpp2020v2xrequirements}.

\begin{table}[t]
\centering
\small
\caption{Parameters for the Munich workload instantiation.}
\label{tab:configuration}
\begin{tabular}{@{}ll@{}}
\toprule
\textbf{Parameter} & \textbf{Value} \\
\midrule
Autonomous vehicle share & 100\% \\
Camera & 13, 10 Hz, 200 KiB/sample \\
LiDAR & 4, 10 Hz, 300 KiB/sample \\
Radar & 6, 10 Hz, 2 KiB/sample \\
Telemetry/raw-data fraction & Camera: 0.002 \\
 & LiDAR: 0.001; Radar: 0.003 \\
Telemetry rate & 0.527 Mbps/vehicle \\
Stored sensor window & $T=5$ s \\
Mean selection rate & $\bar{\lambda}=1$ per vehicle per hour \\
Sensor retention & $a_s=1.0$ \\
Encoded/retained-data fraction & $c_s=0.1$ \\
Training data & 18.6 MiB/event \\
Map data item rate & $u=0.05$ items/vehicle/s \\
Map item & $m=50$, $b_M=200$ bytes \\
Supporting evidence rate & 0.0084 Mbps/vehicle \\
\bottomrule
\end{tabular}
\end{table}

\section{Evaluation}
\label{sec:evaluation}

\subsection{Fleet Workload}

To address RQ1, we test how changes in the hourly number of training-data collection events affect peak workload, using the fixed-rate baseline for comparison. Let $\mu_h=N_h\bar{\lambda}q_h$ denote the expected number of events across the fleet in hour $h$. The Poisson model assumes independent collection events; the variance of the hourly event count equals its expected value, $\mu_h$. To allow larger differences in event counts between repeated runs, we multiply $\mu_h$ by a positive random variable $Z_h$ drawn from a Gamma distribution with shape $k$ and scale $1/k$:
\begin{equation}
Z_h\sim\mathrm{Gamma}(k,1/k),\quad C_h\mid Z_h\sim\mathrm{Poisson}(\mu_hZ_h).
\label{eq:count_model}
\end{equation}
This Gamma--Poisson mixture is a Cox count model~\cite{cox1955series}. Here, $Z_h$ has mean one and variance $1/k$, so $C_h$ has variance $\mu_h+\mu_h^2/k$. A smaller $k$ allows the event count in an hour to vary more between repeated runs. Operational counts of training data selections are unavailable, so $k$ is tested rather than estimated. We use $k=4$ as the main setting and test $k\in\{2,4,8\}$ with 1,000 runs each. In each run, the sampled number of collection events $C_h$ gives an hourly training-data rate of $C_hB_L/3600$.

\begin{figure}[t]
\centering
\includegraphics[width=\linewidth]{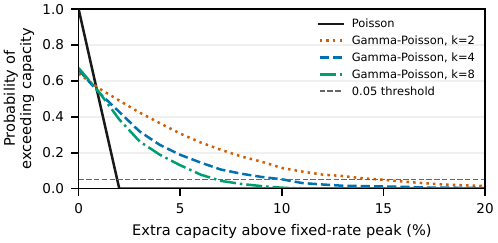}
\caption{Probability that the daily peak exceeds the fixed-rate baseline peak plus the extra capacity shown on the horizontal axis.}
\label{fig:stochastic_workload}
\end{figure}

\begin{table}[t]
\centering
\small
\caption{Daily peak variation under Poisson and Gamma--Poisson counts.}
\label{tab:stochastic_summary}
\begin{tabular}{lrrr}
\toprule
\textbf{Model} & $\mathbf{Var(Z_h)}$ & \textbf{95th-percentile} & \textbf{Extra capacity} \\
 & & \textbf{peak (Mbps)} & \textbf{for $P\leq0.05$} \\
\midrule
Poisson & 0 & 341.9 & 2\% \\
$k=2$ & 0.500 & 384.9 & 15\% \\
$k=4$ & 0.250 & 370.5 & 11\% \\
$k=8$ & 0.125 & 359.7 & 7\% \\
\bottomrule
\end{tabular}
\end{table}

Fig.~\ref{fig:stochastic_workload} shows how often the daily peak requires a higher upload rate than the available capacity. In Table~\ref{tab:stochastic_summary}, 95\% of runs have a peak at or below the reported value. Extra capacity is the smallest tested increase above the baseline that keeps the probability $P$ of exceeding capacity at or below 0.05. Poisson counts require 2\% extra capacity. For $k=4$, the probability falls from 0.661 when capacity equals the fixed-rate baseline peak to 0.188, 0.052, and 0.031 with 5\%, 10\%, and 11\% extra capacity, respectively.

\subsection{Effect of the Hourly Weights}

To address RQ1, we test how the timing of training-data collection affects peak workload while keeping the daily training-data volume unchanged. We move the hourly weights forward by 0--23 hours, wrapping around midnight. After each shift, we adjust all weights by the same factor so that the total training-data volume stays unchanged. Vehicle counts remain unchanged.

\begin{figure}[t]
\centering
\includegraphics[width=\linewidth]{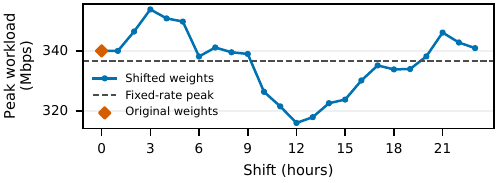}
\caption{Peak workload after shifting the hourly weights while preserving daily training-data volume. A zero shift represents the original timing of training-data collection.}
\label{fig:temporal_alignment}
\end{figure}

Fig.~\ref{fig:temporal_alignment} gives peaks of 315.9--353.7 Mbps at the same 178.4 Mbps daily mean; 14 shifts exceed the fixed-rate baseline peak. With the original hourly weights, the peak is 339.9 Mbps. Peak workload is higher when more vehicles are active during hours with more collection events per vehicle.

\subsection{Parameter Sensitivity}

To address the workload-parameter part of RQ2, we vary each parameter from 0.25 to four times its value in Table~\ref{tab:configuration}, keeping the other parameters unchanged. These factors correspond to telemetry rates of 0.132--2.108 Mbps per vehicle, training data selection rates of 0.25--4 per vehicle per hour, and map evidence rates of 0.0021--0.0336 Mbps per vehicle.

\begin{figure}[t]
\centering
\includegraphics[width=\linewidth]{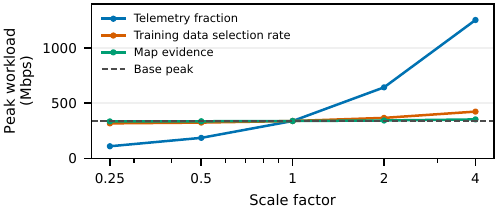}
\caption{Sensitivity of peak fleet workload to the telemetry fraction, training data selection rate, and map evidence rate.}
\label{fig:sensitivity}
\end{figure}

Fig.~\ref{fig:sensitivity} shows that doubling the telemetry fraction, training data selection rate, or map evidence rate raises the peak from 339.9 Mbps to 644.3, 368.3, or 344.8 Mbps, respectively. Their shares of mean demand are 90.4\%, 7.4\%, and 2.1\%. Telemetry therefore determines sustained capacity, while training data affect the peak more than map evidence.

\subsection{Scheduling Training Data}

To address the transfer-delay part of RQ2, we test whether delaying training-data uploads reduces peak workload. We test maximum delays $D$ of zero, one, two, and four hours. These are test settings for offline training uploads, not validated operational deadlines or delays for real-time driving. Telemetry and map uploads remain unchanged. All training data must be uploaded within $D$ hours of collection and before the simulated day ends. We test different capacity limits using binary search and retain the lowest limit that meets every deadline. At each hour, data with the earliest deadline are uploaded first (EDF)~\cite{liu1973scheduling}.

\begin{table}[t]
\centering
\small
\caption{Peak reduction from scheduling training data.}
\label{tab:scheduling}
\begin{tabular}{lrrrr}
\toprule
$D$ (h) & 0 & 1 & 2 & 4 \\
\midrule
Peak (Mbps) & 339.9 & 319.7 & 311.6 & 311.6 \\
Reduction (\%) & 0.0 & 5.9 & 8.3 & 8.3 \\
\bottomrule
\end{tabular}
\end{table}

Table~\ref{tab:scheduling} shows peak reductions of 5.9\% and 8.3\% when uploads can be delayed by one and two hours, respectively. Longer delays allow training data to be uploaded during less busy hours instead of adding to the peak. Beyond two hours, the peak stays at 311.6 Mbps because telemetry and map uploads alone reach this rate and are not delayed. No training data are discarded. The fixed-rate baseline does not separate training data from other uploads, so it cannot specify which data can be delayed.

\subsection{Access Point Capacity Evaluation}

To address RQ3, we compare equal capacity allocation with allocation based on earlier workload, keeping total capacity equal. We use 109 OpenCelliD locations within central Munich as candidate access points~\cite{opencellid}. Each vehicle connects to the nearest access point. We use eight hours of workload to set capacity at each access point, then keep that capacity fixed during the next four hours. The first test uses hours 0--7 to set capacity and hours 8--11 to evaluate it. We repeat this five times, moving the start forward by two hours each time. Capacity is always set using earlier hours, not the hours being evaluated.

Equal allocation divides total capacity equally among access points. Demand-based allocation divides 30\% equally and assigns the remaining 70\% according to the mean workload at each access point during the preceding eight hours. Busier access points therefore receive more capacity, while every access point receives a minimum share. The 30/70 split is a fixed test setting. For both methods, a capacity factor of one means total capacity equals the mean fleet workload during those eight hours; a factor of two doubles that capacity.

For each access point and evaluation hour, we count the upload demand above its assigned capacity as unserved. We sum this excess across access points and hours and divide by total upload demand. We compare the capacity needed to keep this unserved share at or below 5\%; this is a comparison threshold, not a service guarantee.

\begin{figure*}[t]
\centering
\includegraphics[width=\linewidth]{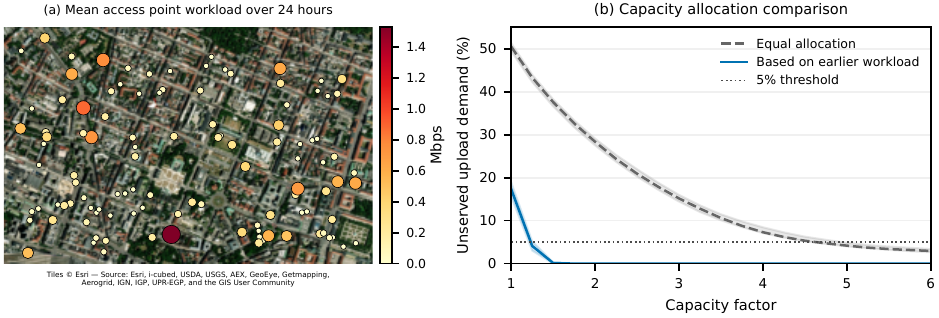}
\caption{(a) Mean access point upload workload over 24 hours, including inactive hours; larger, darker markers indicate higher workload. (b) Unserved demand under equal total capacity. Lines show medians across five tests; bands show ranges.}
\label{fig:edge_provisioning}
\end{figure*}

Fig.~\ref{fig:edge_provisioning}(a) shows how mean upload workload differs across access point locations. Fig.~\ref{fig:edge_provisioning}(b) shows that, at twice the earlier mean fleet workload, demand-based allocation leaves a median of 0\% of demand unserved, compared with 28.4\% under equal allocation. To keep the unserved share at or below 5\%, the median required capacity factor is 1.25 for demand-based allocation and 4.75 for equal allocation. These results answer RQ3: using earlier workload to distribute the same total capacity reduces unmet upload demand in this scenario.

\subsection{Use in Communication and Edge Experiments}

The generator exports workload traces for vehicles, the fleet, and access points, including runs with randomly sampled collection events. These traces support scheduling, placement, and allocation experiments~\cite{wang2020online,chen2023mobility}. The code reproduces results and checks data-volume conservation and equal total capacity.

%% file: sections/05_conclusion.tex
\section{Conclusion and Future Work}
\label{sec:conclusion}

We introduced an autonomous vehicle offloading workload generator that combines application requirements with vehicle mobility. It models telemetry, training data, and HD map observations separately and reports workload for individual vehicles, the fleet, and each access point.

In the Munich case study, changing when training data are collected changes peak workload even when daily data volume stays the same. Across the tested timing shifts, the highest peak is about 12\% higher than the lowest. Larger fluctuations in the hourly number of collection events require more capacity to keep the probability of exceeding capacity low.

Peak workload is most sensitive to the telemetry fraction in the tested configuration. The model separates training data from telemetry and map data, so training-data uploads can be delayed without delaying the other uploads. With a two-hour deadline, earliest deadline first scheduling reduces the peak by 8.3\% without discarding data. With the same total capacity set to twice the earlier mean fleet workload, demand-based allocation reduces the median unserved share from 28.4\% under equal allocation to 0\%.

The generator provides workload as input to systems that model radio conditions, queuing, handover, and computation. Future work will replace the hourly weights derived from accident records with operational records of training data selection, evaluate other cities, and combine the workload with network and vehicle models to study whether communication and processing delays leave enough time for emergency braking.